\documentclass[preprint,aps, preprintnumbers, nofootinbib]{revtex4}
\usepackage[dvips]{graphics}

\newcommand{\lsim}[1]{
\setlength{\unitlength}{12pt}
\begin{picture}(1.4,1.)
\put(.7,-0.3){\makebox(0.0,1.)[t]{$<$}}
\put(.7,-0.3){\makebox(0.0,1.)[b]{$\sim$}}
\end{picture}#1}

\newcommand{\gsim}[1]{
\setlength{\unitlength}{12pt}
\begin{picture}(1.4,1.)
\put(.7,-0.3){\makebox(0.0,1.)[t]{$>$}}
\put(.7,-0.3){\makebox(0.0,1.)[b]{$\sim$}}
\end{picture}#1}

\begin{document}

\title{Avoiding Boltzmann Brain domination in holographic dark energy
models}

\author{R. Horvat}
\email{horvat@irb.hr}
\address{Rudjer Bo\v{s}kovi\'{c} Institute, P.O.B. 180, 10002 Zagreb,
Croatia}

\begin{abstract}
In a spatially infinite and eternal universe approaching ultimately
a de Sitter (or quasi-de Sitter) regime, structure can form by thermal 
fluctuations as such a space is thermal. The models of Dark Energy invoking
holographic principle fit naturally into such a category, and spontaneous
formation of isolated brains in otherwise empty space seems the most
perplexing, creating the paradox of Boltzmann Brains (BB). It is thus
appropriate to ask if such models can be made free from domination by Boltzmann
Brains. Here we consider only the simplest model, but adopt both the local
and the global viewpoint in the description of the Universe. In the former case, we find that if a dimensionless model 
parameter $c$,
which modulates the Dark Energy density, lies outside the exponentially narrow strip
around the most natural $c = 1$ line, the theory is rendered BB-safe.
In the later case, the bound on $c$ is exponentially stronger, and seemingly
at odds with those bounds on $c$ obtained from various observational tests.

\end{abstract}

%\pacs{97.60.Gb, 14.60 Pq, 04.80.Cc}
\newpage

\maketitle

An empty space with positive cosmological constant (attainable after all
kinds of matter are emptied out) represents a thermal system, having a non-zero
temperature as well as the maximal entropy \cite{1}. At rare occasion such a system (if
sufficiently long-lived) would spontaneously form structures as a thermal
fluctuation. This assumes a downward shift in entropy, and it is just this drift
from the generalized second law of thermodynamics \cite{2, 3, 4} that makes virtually
any pop-up structure devoid of having a conventional history record
\cite{5}. Namely, amongst all observers created spontaneously out of a
thermal system the vast majority of them correspond to the smallest
fluctuation - isolated brains immersed in thermal equilibrium of the empty
space. This constitutes the paradox of Boltzmann Brains (BBs) \cite{6, 7, 8} - when ordinary observers
(related to the conventional formation and evolution of structures via inflation and subsequent
reheating of the early Universe) become vastly outnumbered by those who
(having the same impressions and the same frame of mind)
form spontaneously out of a sufficiently long-lived vacuum. So it is
appealing to see if there is an escape for any  otherwise  viable cosmological theory from this
troublesome situation. 

If the vacuum decays fast enough into a different vacuum, the undecayed
physical volume then stops growing before the  production of BBs
is initiated \cite{8}. For our universe the decay time can be calculated to
be of order $10^{10}$
yr \cite{8}. This resolution of the BB paradox, however, poses a serious problem for a 
description of the
multiverse if the global viewpoint in the description of the Universe 
is adopted \cite{9}. Recently, it was
noticed \cite{10} that the BB threat is not specific only for those exotic theories like the
string theory multiverse, but such an unpleasant situation may be found even
in the vanilla $\Lambda$CDM model, in relation with the electroweak vacuum
and ordinary physics. It is interesting to note  that without new physics,
the Page's resolution for the electroweak vacuum 
works only if the top pole mass lies somewhat beyond the current observational
bounds \cite{10}. Also, the role of phantom cosmologies in treating the BB
paradox was stressed 
recently \cite{11}. 

In the present paper, we consider how holographic dark energy (HDE) models
\cite{12, 13, 14},
as viable set of models for a description of the Dark Energy in the
late-time Universe, cope with the intimidation of the BB brains. The form of
the vacuum energy in HDE models stems from the holographic principle [15, 16],
undoubtedly the most amazing ingredient of a modern
view of space and time. The fate of the Universe in these models proves
notably
susceptible to a slight variation in the vacuum energy density \cite{13}, allowing 
behavior not only similar to the cosmological constant, but the phantom case as
well. The scenario therefore proves susceptible to the BB domination.     

To incorporate the holographic principle in an effective QFT 
necessarily requires a kind
of UV/IR mixing \cite{17}. This is so since  in QFTs the entropy $S \sim L^3 \Lambda^3 $ (where $L$
is the size of the region and $\Lambda$ is the UV cutoff) scales extensively,
and therefore  there is always a sufficiently large volume (for
any  $\Lambda$) for which
$S$ would exceed the absolute Bekenstein-Hawking bound ($\sim M_{Pl}^2 L^2 $). After discarding a
great deal of states with Schwarzschild radius much larger than the box size
(not describable within QFTs), the bound gets more stringent
$(M_{Pl}^2 L^2)^{(3/4)})$ \cite{18}. Near saturation, the bound gives the
following vacuum energy density,
\begin{equation}
\rho_{\Lambda} = (3/8\pi) c^2 M_{P}^2 L^{-2} \;,
\end{equation}
where $c$ is a free parameter introduced in \cite{13}, with a natural
value of order one. 

For the sake of demonstration, we shall consider the appearance of BB brains
in the simplest (i.e. non-interacting) HDE model \cite{13}, where the event
horizon of the spatially flat Universe $R_h $ was chosen for $L$ \footnote{Since for the
purpose of the present paper we are mostly interested in the future
evolution of the Universe, even this simplest model can represent virtually all the
models having the same choice for $L$.}. The model without any
additional energy component is easily
solvable, yielding for the equation of state \cite{13}
\begin{equation}
\omega = - \frac{1}{3} - \frac{2}{3c}\;,
\end{equation}
and $R_h $ scales with the scale factor as
\begin{equation}
R_h = R_{h0} ~ a^{1 - \frac{1}{c}}\;,
\end{equation}
where the subscript ``0" indicates the present epoch. The fate of the Universe
is strictly dictated by the value of $c$: for $c \geq 1$ the Universe
enters  the (quasi-) de Sitter regime ($c=1$ mimics the cosmological
constant), while $c < 1$ corresponds to the phantom regime.

Let us first adopt the local viewpoint. In our case this means that the
interior of $R_h $ is everything there is (the interior corresponds to 
the causally connected region). For $c=1$, thus $R_h $ being a
constant, the number of
ordinary observers stays finite for any time in the future, while on the
other hand  the number of
BB observers starts to pile up after a typical timescale (exponentially
huge) of order \cite{5}
\begin{equation}
t_{BB} \sim exp(E_{br} R_h ) t_{dyn}\;,
\end{equation}    
where $E_{br}$ is the energy of the BB brain and $t_{dyn}$ is a dynamical
timescale typical for the equilibrium system \footnote{The exact expression
for $t_{dyn}$ is not of relevance here since we are dealing with exponentially
huge figures in front of it. For the sake of rendering our calculation (see below) more
compact and not introducing another parameter, we choose $H_{0}^{-1}$ for
$t_{dyn}$, where H is the Hubble parameter.}. The system described by (4) is
a thermal system revealing the (constant) Gibbons-Hawking temperature given in
terms of the inverse radius of the de Sitter space $T_h = 1/(2\pi R_h)$.
Besides,
it also describes thermal fluctuations obtained from the entropy decrease of
the de Sitter space, when the brain of energy $E_{br}$ is formed in the system of size $L$.
Some extremely rare fluctuation would reproduce our whole visible Universe after
the Poincar$\acute{e}$ recurrence time of order $e^{10^{122}}t_{dyn}$ is passed. So
for $c=1$ the Universe is eternally inflating with a constant $R_{h}$, and
in the absence of the landscape of string theory vacua, or stated simply, in
the absence of any other vacuum
our vacuum can decay to, the BB problem is unavoidable.

More subtle analysis is required if $c > 1$ or $c < 1$, since in either case $R_{h}$
(and therefore the Gibbons-Hawking temperature) is
time-dependent. First we consider the issue of whether thermodynamic
equilibrium is maintained also for a time-dependent temperature $T_h$. To
this end, we
adopt a heuristic criterion for maintaining equilibrium in the form
\begin{equation}
\left |\frac{R_h}{\dot{R}_h} \right| \gsim \; \frac{R_h}{c_{\gamma}} \;,
\end{equation}
that is, departures from de Sitter space should be small enough so that the
l.h.s. of (5) is always larger than the light-crossing
time of the radius $R_h$. In a
two-component flat-space universe $\rho_{\Lambda}$ evolution is governed
by \cite{13}
\begin{equation}
\Omega_{\Lambda }^{'} = \Omega_{\Lambda }^2 ( 1 - \Omega_{\Lambda }) \left
[\frac{1}{\Omega_{\Lambda }} + \frac{2}{c \sqrt{\Omega_{\Lambda }}} \right ]
\;,
\end{equation}
where the prime denotes the derivative with respect to $lna$ and
$\Omega_{\Lambda } = \rho_{\Lambda }/\rho_{crit} $. Combining (6)   
with (5) for the matter case one arrives at
\begin{equation}
\left |\frac{\sqrt{\Omega_{\Lambda }}}{c -\sqrt{ \Omega_{\Lambda }}} \right|
\gsim \; 1 \;.  
\end{equation}
Employing $c$ close to 1 (see below), one obtains $\Omega_{\Lambda } > 1/4$.
Thus, thermodynamic equilibrium was being established somewhere around
the onset of the dark-energy dominated epoch and will stay there for anytime in the
future.  
  
Let us consider first the quintessential $c > 1$ case. To
this end, we need to find an  explicit solution $a(t)$ of the Friedmann
equation of the type
\begin{equation}
\dot{a}(t) = H_0 ~ a^{1/c}(t)\;.
\end{equation}
With the normalization $a(t_{0}) = 1$, one finds an explicit solution
\begin{equation}
a(t) = \left[-H_0 t(\frac{1}{c} - 1) + H_0 t_0(\frac{1}{c} - 1) + 1
\right]^{\frac{c}{c-1}}
\;. 
\end{equation}
Plugging (9) into (3), the central equation to be solved
\begin{equation}
t_{BB} \simeq exp(E_{br} R_{h}(t_{BB})) t_{dyn}\;,
\end{equation}
with $E_{br} \sim 1$ kg and $t_{dyn} \simeq H_{0}^{-1} \simeq t_{0}$, can be
recast in the form
\begin{equation}
\eta_{BB}(c) \simeq exp \left( 10^{68} \left[\eta_{BB}(c) (c - 1) +
1 
\right]
\right)\;,
\end{equation}
where $\eta_{BB} \equiv t_{BB}H_0 $. Notice that if Eq.(11) has no solution
for any $c$, then the theory can be considered BB-free. Because of the
exponentially huge figures entering (11), it is unfortunately extremely
difficult to handle it numerically, and therefore one has to resort to
heuristic methods in order to infer some information on the parameter $c$.
For instance, one can easily solve (11) for $c(\eta_{BB})$, i.e., 
\begin{equation}
c(\eta_{BB}) = \frac{\eta_{BB}A - A + ln(\eta_{BB})}{A\eta_{BB}};,
\end{equation}
where $A = 10^{68}$. By inspecting (12), one sees that for $\eta_{BB} =
exp(A)$ and $\eta_{BB} = \infty$ $c = 1$, and therefore somewhere within
this interval (12) reveals a maximum. With the maximum expressed in a
closed form, we find that we are exposed to the BB threat if 
\begin{equation}
1 < c <  1 + \left(\frac{1}{A}\right)  e^{-A -1}\;.  
\end{equation}
Still, since for $c > 1$ the event horizon $R_h$ grows in
time, the number of ordinary observers is growing with time as well, 
making it hard to reckon the real BB threat. All we can say for sure is that
if the parameter $c$ lies outside the exponentially narrow strip right to
the $c
= 1$ line, as given by (13), the theory is safe with respect to the BB invasion.
 
Let us next consider the phantom regime ($c < 1$). Now the scale factor (9)
diverges after finite time is passed - the big rip time \cite{19}. Also, for $c
< 1$ $R_h$ decreases in time, falling to zero at the big rip time. The
potential BB threat lasts until $R_{h}^{smallest} \sim 10$ cm is reached, the
smallest possible size of the event horizon capable of housing a single BB
observer \footnote{Notice that decreasing $R_h$ leads to a much higher 
BB creation rate, see Eq. (10).}. Thus, if $t_{smallest} \lsim t_{BB}$, then BB brains are avoided.
This means that in addition to (12), an extra constraint
\begin{equation}
1 + \frac{c}{1-c} \left(1 - \frac{B}{c}\right) \lsim \eta_{BB}(c) \;,
\end{equation}
is to be considered in the phantom regime. In (14) $1/B = 10^{27}$. Now a
straightforward analysis of (12), together with the constraint (14), enables one to
obtain in a closed form  an interval where one is to be exposed to the BB threat,
i.e.,
\begin{equation}
1 > c > 1 + e^{-\frac{A}{B}} (B -1) 
\end{equation}

Our analysis thus shows that only in the exponentially narrow strip around the $c
= 1$ line,
\begin{equation}
1 + e^{-\frac{A}{B}} (B -1) < c < 1 + \left(\frac{1}{A}\right)  e^{-A -1}\;,
\end{equation}
problems with BBs in the simplest HDE model are to be expected. We notice
that a restriction on the parameter $c$ under the combined observational
tests slightly favor the phantom case \cite{20, 21, 22, 23}. 

Finally, let us adopt the global viewpoint. This means that 
exponentially huge regions created by the expansion of the Universe, which
no one observer can ever probe, are also accepted as a part of reality.  
As already mentioned, a description 
of the BB paradox in the global picture leads to strong inconsistencies,
if the idea of the string theory landscape is also to solve the cosmological
constant problem \cite{9}. Likewise, the popular resolution of the black
hole information paradox is to abandon the global viewpoint, and to embrace the
local view in the form of the black hole complementarity principle
\cite{24}. For these reasons we give much more preference for the local
view, but for completeness sake we analyze the global viewpoint as well. To
this end, we wholly follow  the Page's arguments \cite{7}.

Following Page \cite{7}, when a four-volume of the Universe exceeds $
V_{4}^{crit} \sim e^{10^{50}} a_{\rm Pl}^4 $, one gets more observations by vacuum
fluctuation than have occurred during past human history. Specifically,
\begin{equation}
V_{4}(t) = \int d^4 x \sqrt{-g} ~ \sim ~ \int_{t_0 }^{t} dt a^3 (t) \;, 
\end{equation}
with $a(t)$ from (6), as obtained in the HDE model. Obviously, if $c \geq 1$,
the four-volume (17) grows unlimited  with time, and without the vacuum
decay, the BB problem persists.

Much more interesting is the phantom $c < 1$ case. Now the Universe lasts
only until the Big Rip time
\begin{equation}
t_{BR} = t_0 + \frac{c}{(1-c)H_0 } \;,
\end{equation}
where the scale factor (9) becomes infinite. It can be seen by inspecting
Eq.(17)
that in this case the four-volume can be made finite for $c < 1/4 $. At the same time 
the requirement $V_{4} \lsim V_{4}^{crit}$ provides us with a bound
on the parameter $c$
\begin{equation}
c < \frac{1}{4} \left(1 - \frac{1}{4C} \right)\;,
\end{equation}  
where $C =  (H_0 l_{Pl})^4 e^{10^{50}}$.
This is exponentially  more restrictive then the bound obtained by adopting
the local view. On the
other hand, the
bound $c < 1/4 $ seems at odds with those bounds obtained from a variety of
observational tests \cite{20, 21, 22, 23}. To be specific, focusing on the most thorough 
study \cite{20}, we see that the best fit of $c$ for the simplest model
\cite{13} gets centered around $0.75$, for all data set combinations used in
this study. Although there is some fraction of allowed parameter space in
which $c > 1$, there is absolutely no room for $c$ as low as $1/4$. Also,
other studies \cite{21, 22, 23} using less precise data gave the best fit value
even higher than $0.75$, thus moving away even more from the value $1/4$. So
the observational signatures of the simplest HDE model yet 
additionally apostrophize the known difficulties of the global view.

Summing up, we have tested how the simplest holographic dark energy model of Li
copes with the theoretical conundrum known as the Boltzmann Brain paradox.
And the outcome strongly depends on the description of the Universe
(whether local or global) one adopts. With the local viewpoint, there is no
restriction (up to an exponentially negligible one) on the free parameter of the
theory, whilst the global viewpoint sets a  restriction on it much stronger than
those obtained from observational tests.  In absence of any theoretical
constraint on the energy density parameter $c$, our constraints may be
considered as a new and useful piece of information corroborating further
the genuine quantum-gravity origin of the model.

{\bf Acknowledgment. } This work has been fully supported  by Croatian  
Science Foundation under the project (IP-2014-09-9582).

\end{document}